\documentclass[aps,prl,showpacs,notitlepage,reprint]{revtex4-1}
\usepackage{graphicx}
\usepackage{dcolumn}
\usepackage{amsmath}
\usepackage{bm}
\usepackage{todonotes}
\usepackage[utf8]{inputenc}
\usepackage{amsmath}
\usepackage{natbib}
\usepackage{graphicx}
\usepackage{caption}
\usepackage{subcaption}
\usepackage{lipsum}
\usepackage{float}

\newcommand{\bn}[1]{\mbox{\boldmath$#1$}}

\newcommand{\beq}{\begin{equation}}
\newcommand{\eeq}{\end {equation}}
\newcommand{\bea}{\begin{eqnarray}}
\newcommand{\eea}{\end{eqnarray}}

\begin{document}
\title{Super-chirality of paraxial higher order Poincar{\'e} modes} 
\author{{\rm {M. Babiker}}$^{1,*}$, J. Yuan$^{1}$, K. Koksal$^{2}$, V. E. Lembessis$^{3}$} 
\affiliation{$^1$School of Physics, Engineering and Technology, University of York, YO10 5DD, UK}
\affiliation{$^2$Physics Department, Bitlis Eren University, Bitlis, Turkey}
\affiliation{$^3$Quantum Technology Group, Department of Physics and Astronomy, College of Science,
King Saud University, Riyadh 11451, Saudi Arabia}
\vspace{10mm}
\affiliation{$^*$ Corresponding author: m.babiker@york.ac.uk}
\date{\today}

\begin{abstract}
We demonstrate that higher order Poincar{\'e} modes of order $m$ are super-chiral, displaying enhancement factors proportional to $m$ and  $m^2$ in their helicity/chirality. With $m$ having arbitrarily large integer values, such modes, in  principle, possess unlimited super-chirality. These findings pave the way to applications, including the strong enhancements of optical interactions with chiral matter. The work indicates considerable flexibility in  controlling the helicity of any higher order paraxial twisted light mode and it incorporates a very wide range of physical scenarios.
\end{abstract}

\maketitle

Recent research has highlighted the fundamental significance and the potential for applications of higher-order optical vector modes, also called Poincar{\'e} modes \cite{milione2011,GALVEZ202195,maurer2007,liu2014,naidoo2016,Zhan2009,volpe2004,Holmes_2019}.
The overall polarisation state of such modes is formally identified as characteristically non-separable superpositions of solutions involving  circular polarisation $({\bn {\hat x}}\pm i{\bn {\hat y}})/\sqrt{2}$ and spatial phase functions $e^{\pm im\phi}$ with integer $m$ the higher order and $\phi$ the azimuthal angular variable.
The  polarisation is represented by a point $(\Theta_P,\Phi_P)$ on the surface of a unit Poincar{\'e} sphere, as explained in the caption to Fig.1.  


Although the higher order vector modes have already been realised experimentally \cite{liu2014,naidoo2016,chen2020}, their properties 
have not yet been  explored for arbitrary $m$.  In particular,  the question arises as to whether and how the higher order modes can offer enhanced beam properties such as higher order encoding schemes for enhanced bandwidth optical communications \cite{al2021structured}  and whether they could lead to enhanced optical angular momentum, spin and chirality which could influence optical interaction with chiral matter \cite{Tang2010}. Increased optical chirality is highly desirable in order to engage effectively with chiropical processes. Could it be the case that higher order modes would provide sufficiently strong chirality to engage effectively with chiral molecules and be able to achieve a high degree of enantioselectivity?

In this Letter we focus on the prospect  of the existence of super-chirality, which,  we envisage, maybe one of the major properties of the higher order modes. To this end, we have aimed to evaluate the helicity density and its spatial integral for the most general paraxial  mode of arbitrary order $m\geq 0 $, which covers all possible scenarios.  

In cylindrical coordinates the electric and magnetic fields of a general paraxial twisted light mode with the most general polarisation are derivable from a vector potential in the form
\beq
{\bf A}={\bn {\hat \epsilon}}{\tilde {\cal F}}_{\{\ell\}}(\rho)e^{i\ell\phi}e^{ik_zz}
\label{vect1}
\eeq
Here $k_z$ is the axial wavevector with the light travelling along the $+z$ axis and ${\tilde {\cal F}}_{\{\ell\}}$ is the paraxial mode function which depends only on the radial coordinate $\rho$. The mode is  labelled by the group of indices generically denoted by $\{\ell\}$, which includes integer $\ell$, the winding number, and integer $p$ the radial number, as in the case of Laguerre-Gaussian (LG) optical vortex modes.  However, the treatment is not restricted to LG modes and is applicable, in general, to other vortex modes.  
The higher order polarisation state vector can be written as
\beq
{\bn {\hat \epsilon}}=e^{im\phi}({\bn {\hat x}}-i{\bn {\hat y}}){\cal U}_P+e^{-im\phi}({\bn {\hat x}}+
i{\bn {\hat y}}){\cal V}_P
\eeq
where $m$ is a positive integer, unlike $\ell$ which spans all real integers; ${\cal U}_P$ and ${\cal V}_P$ are Poincar\'e functions given by
\beq 
 {\cal U}_P=\frac{1}{\sqrt{2}}\cos{\left(\frac{\Theta_P}{2}\right)}e^{-i\Phi_P/2}
 ;\;\;\;\;{\cal V}_P=\frac{1}{\sqrt{2}}\sin{\left(\frac{\Theta_P}{2}\right)}e^{i\Phi_P/2}
 \label{upee}
\eeq
The above polarisation state ${\bn {\hat \epsilon}}$ is the most general polarisation vector, similarly defined by Milione et al \cite{milione2011} in terms of the higher order Poincar\'e sphere. The validity of the higher order polarisation states, has already been confirmed experimentally \cite{liu2014,naidoo2016,shen2020}. 

At a general point on the surface of the higher order Poincar\'e sphere, we have for the vector potential 
\begin{widetext}
\beq
{\bf A}=\left\{({\bn {\hat x}}-i{\bn {\hat y}})e^{i(\ell+m))\phi}{\cal U}_P+({\bn {\hat x}}+i{\bn {\hat y}})e^{i(\ell-m)\phi}{\cal V}_P\right\}{\tilde {\cal F}}_{\{\ell\}}(\rho)e^{ik_zz}
\label{vect2}
\eeq
\begin{figure}
\includegraphics[width=0.8\columnwidth]{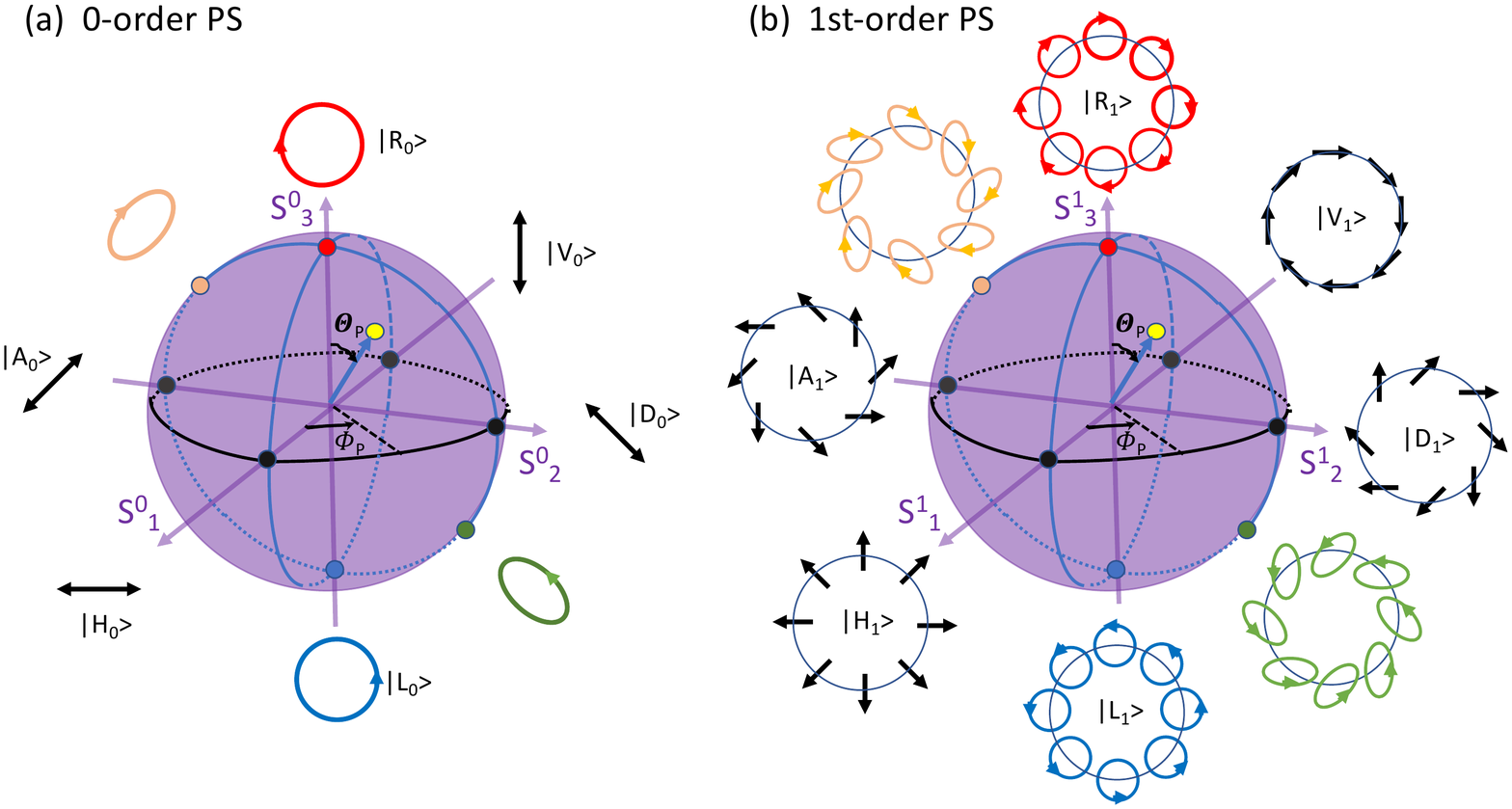}
\caption{0th order, (a), and 1st-order, (b), Pioncar\'e Sphere (PS) representation of the  polarisation state in which optical polarisation is coupled with vortex phase, characterized by a unit sphere with respect to the corresponding Stokes-parameter(-like) Cartesian coordinates ($S^0_1, S^0_2$, $S^0_3$) and ($S^1_1, S^1_2$, $S^1_3$) respectively.  It is seen that the 0th order PS is equivalent to the conventional PS where $|H>$ and $|V>$ are commonly used to denote the vertically and horizontally linearly polarized light, $|A>$ and $|D>$ for $\pm 45^o$ tilted linearly polarized light, $|R>$ and $|L>$ for right-hand and left-hand circularly-polarized light, respectively.   The 1st order PS figure is related to the corresponding figure by Milione et al \cite{milione2011} with slightly different conventions for $S^1_1$ and $S^1_2$.  Six sets of special vector modes are drawn in different colours next to each sphere for illustration.  Their positions on the Poincar\'e sphere are indicated by dots of the same colour.}
\end{figure}
\end{widetext}
An important requirement of free-space paraxial optical fields is that the electric field must be derivable from the magnetic field using the first Maxwell curl equation and that the electric field must produce the same magnetic field via the second Maxwell curl equation. We  write the vector potential as the sum of two terms 
\bea
{\bf A}&=&{\bf A_1}+{\bf A_2}\label{combvec}\\
{\bf A_1}&=&({\bn {\hat x}}-i{\bn {\hat y}}){\cal F}^{(1)}_{\{\ell\}}(\rho,\phi)e^{ik_zz}\\
{\bf A_2}&=&({\bn {\hat x}}+i{\bn {\hat y}}){\cal F}^{(2)}_{\{\ell\}}(\rho,\phi)e^{ik_zz}\label{vect3}
\eea
where we have introduced ${\cal F}^{(i)}_{\{\ell\}}$ with $i=1,2$ as the functions of $\rho$ and $\phi$ as follows
\bea
{\cal F}^{(1)}_{\{\ell\}}(\rho,\phi)&=&{\cal U}_P e^{i(\ell+m)\phi}{\tilde {\cal F}}_{\{\ell\}}(\rho);
\nonumber\\
{\cal F}^{(2)}_{\{\ell\}}(\rho,\phi)&=&{\cal V}_P e^{i(\ell-m)\phi}{\tilde {\cal F}}_{\{\ell\}}(\rho)
\label{eff1}
\eea

The electric and magnetic fields of this generally-polarised mode are similarly written as the sums ${\bf B}={\bf B}_1+{\bf B}_2$ and ${\bf E}={\bf E}_1+{\bf E}_2$ where ${\bf B}_i={\bn {\nabla}}\times{\bf A}_i;\;\;\;i=1,2$. The sequence of steps involve dealing first with the two parts of the magnetic field and from those use Maxwell's curl {\bf B} equation to derive the corresponding electric field parts.  We have for ${\bf B}_1$ and ${\bf E}_1$ 
\bea
&&{\bf B}_1=\left\{ik_z({\bn {\hat y}}+i{\bn {\hat x}})-{\bn {\hat z}}\left(i\partial_x +\partial_y\right)\right\} {\cal F}^{(1)}e^{ik_zz}\nonumber\\
&&{\bf E}_1=c\left\{ik_z({\bn {\hat x}}-i{\bn {\hat y}})-{\bn {\hat z}}\left(\partial_x -i\partial_y\right)\right\} {\cal F}^{(1)}e^{ik_zz}
\label{emfields1}
\eea

It is easy to see that ${\bf B}_2$ and ${\bf E}_2$ follow, respectively, from ${\bf B}_1$ and ${\bf E}_1$ by the following substitution
\bea
{\bf B}_2&=&{\bf B}_1(i\rightarrow-i;{\cal F}^{(1)}e^{ik_zz}\rightarrow {\cal F}^{(2)}e^{ik_zz})\nonumber\\
{\bf E}_2&=&{\bf E}_1(i\rightarrow-i;{\cal F}^{(1)}e^{ik_zz}\rightarrow {\cal F}^{(2)}e^{ik_zz})
\label{emfields2}
\eea
where we have dropped the subscript label $\{\ell\}$ in ${\cal F}^{(1),(2)}$ and in ${\tilde {\cal F}}$ for ease of notation and the notation can be restored when the need arises. It is easy to see that the procedure we have followed amounts to ensuring that the fields satisfy the wave equation ${\bn {\nabla}}\times{\bn {\nabla}}\times{\bf E}-\omega^2{\bf E}/c^2=0$ to the leading derivative order. The fields we now have form the basis for the derivation of the optical properties of the higher order modes.

The cycle-averaged optical densities of the  helicity ${\bar \eta}$ and chirality ${\bar \chi}$  are defined by
\beq
{\bar \eta}({\bf r})=-\frac{\epsilon_0 c}{2\omega}\Im\sum_{i=1}^2\sum_{j=1}^2\left({\bf E}_i^*\cdot{\bf B}_j\right)=\frac{c}{\omega^2}{\bar \chi}
\label{chaid}
\eeq
where $\omega$ is the frequency of the light, ${\bf E}_{i}$ and ${\bf B}_{i}$, with $i=1,2$ are as given in Eqs.(\ref{emfields1}) and (\ref{emfields2}).The symbol $\Im[...]$ in Eq.(\ref{chaid}) stands for the imaginary part of [...] and the superscript * in ${\bf E}^*$ stands for the complex conjugate of ${\bf E}$.  In what follows, we focus on the helicity from which the chirality can be determined using  Eq.(\ref{chaid}). 

We seek to evaluate both the helicity density and its space integral specifically in relation to the most general higher order optical vortex modes.
The four terms arising from the summation in Eq.(\ref{chaid}) require separate evaluations. The evaluations are straightforward and require, as a first step, expressions for the  x- and y-derivatives of ${\cal F}^{(1)}$ and ${\cal F}^{(2)}$ in polar coordinates.  Note from Eqs.(\ref{eff1}) that ${\cal F}^{(1)}$ is distinguished by the phase factor $\exp{[i(\ell+m)\phi]}$ and ${\cal F}^{(2)}$ is distinguished by the phase factor $\exp{[i(\ell-m)\phi]}$. 
It turns out that the sum  ${\bf E}_1^*\cdot{\bf B}_2+{\bf E}_2^*\cdot{\bf B}_1$ does not contribute an imaginary part and only the two direct terms contribute.  We find after some algebra
\begin{widetext}
\beq
 {\bar {\eta}}({\bf r})=\frac{\epsilon_0 c^2}{4\omega}\left\{\cos{(\Theta_P)}\left[\left(2k_z^2|{\tilde{\cal F}}|^2+|{\tilde{\cal F}'}|^2\right)+\frac{\ell^2}{\rho^2}|{\tilde {\cal F}}|^2\right]-2\ell\frac{{\tilde {\cal F}'}{\tilde{\cal F}}}{\rho}\right\}
 +\frac{\epsilon_0 c^2}{4\omega}\left(\frac{|{\tilde{\cal F}}|^2}{\rho^2}[m^2\cos{(\Theta_P)}-2m\ell]+\frac{{\tilde {\cal F}'}{\tilde{\cal F}}}{\rho}m\cos{(\Theta_P)}\right)
 \label{heldens3}
\eeq
\end{widetext}
where we have set ${\tilde{\cal F}'}=d{\tilde{\cal F}}/d\rho$ and chosen to separate the $m$-dependent terms from the other terms.  The first set of terms in Eq.(\ref{heldens3}) (enclosed between the curly brackets) coincides with the  zero order ($m=0$) helicity density in the case of elliptical polarisation.  The rest are the $m$-dependent higher-order terms and are capable for sufficiently large $m$ of dominating the zero-order helicity, leading to super-chirality. 
The Poincar\'{e} function $\cos(\Theta_P)$ takes real values from +1.0 ($\Theta_P=0$; right-hand circular polarisation at the north pole of the Poincar\'{e} sphere) to -1.0 ($\Theta_P=\pi$; left-hand circular polarisation at the south pole) with intermediate points $0<\Theta_P<\pi$ representing elliptical polarisation and the special points where $\Theta_P=\pi/2$ representing radial and azimuthal polarisation. 
Thus we can immediately infer that we have a very general result which is applicable to any paraxial  optical vortex of a general polarisation defined by a point on the surface of a higher order Poincare sphere.  To obtain the helicity density for any specific case all we need to do is simply specify the order $m$, the Poincar\'{e} polarisation angles $(\Theta_P, \Phi_P)$ and the amplitude function ${\tilde {\cal F}}$, with its winding number $\ell$ and its radial number $p$, if applicable. Note, however, that the helicity does not contain any dependence on the Poincar\'{e} angle $\Phi_P$, so that, for example, all points on the equatorial circle have the same helicity.

Setting $m=0$ and $\Theta_P=0,\pi$ in Eq.(\ref{heldens3}) we immediately identify the exact expression between the curly brackets as the helicity density of the basic circularly-polarised general optical vortex mode \cite{babiker2022}, interpreting $\cos(\Theta_P)=\pm 1$ as $\sigma=\pm 1$ for circular polarisation.  There is also an additional term involving $-\ell\frac{{\tilde {\cal F}'}{{\tilde{\cal F}}}}{\rho}$, which is appropriate for uniform linear polarisation and has been shown to lead to zero chirality  on spatial integration \cite{babiker2022,koksal2022e,forbesb2021,forbes2022}. The basic circularly-polarised helicity defined by the terms in the curly brackets has been fully evaluated for Laguerre-Gaussian light \cite{koksal2022e}.

However, for $m\neq 0$ there are now additional $m$-dependent terms in the helicity density for all values of $\cos{(\Theta_P)}=(+1.0\; {\rm {to}}\; -1.0)$, which means that elliptically polarised modes (including circular,  linear, as well as radial and azimuthal) have additional $m$-dependent density contributions.  In particular, 
For $m\geq 1$, as $\Theta_P$ increases the Poincar\'{e} function $\cos(\Theta_P)$ passes through zero at $\Theta_P=\pi/2$  for all points $\Phi_P$ on the equatorial circle. Only for $m=1$, this helicity density coincides with the case of radially-polarised optical vortex modes \cite{koksal2022d}. We have for $m>1$
\beq  
{\bar {\eta}}_{m\ell}({\bf r})=-\ell\frac{\epsilon_0 c^2}{2\omega}\left\{m\frac{|{\tilde{\cal F}}_{\{\ell\}}|^2}{\rho^2}+\frac{{\tilde {\cal F}'}_{\{\ell\}}{{\tilde{\cal F}}}_{\{\ell\}}}{\rho}\right\}\label{emm}
\eeq
Clearly, since $m$ can in principle take any integer value greater than 1, the first term in Eq.(\ref{emm}) increases with increasing $m$. This means that the magnitude of the term is $m$ times larger than for the case $m=1$, which corresponds to lowest order radially-polarised paraxial modes. 
The general case for which $m>1$ and for any point ${\Theta_P,\Phi_P}$, the helicity density is given by Eq.(\ref{heldens3}) and it constitutes the most general result for the helicity density of a paraxial vortex mode of any order $m$. 

We may now evaluate the super-chirality properties of higher order modes for the special case of a paraxial Laguerre-Gaussian mode of winding number $\ell$, radial number $p$ and waist $w_0$, which has an amplitude function given by
\beq
{\tilde {\cal F}}_{\ell,p}(\rho)={\cal E}_0\sqrt{\frac{p!}{(p+|\ell|)!}} e^{-\frac{\rho^2}{w_0^2}}  \left(\frac{\sqrt{2}\rho}{w_0}\right)^{|\ell| }L^{|\ell|}_p\left(\frac{2\rho^2}{w_0^2}\right)\label{efftilde}
\eeq
where $L_p^{|\ell|}$ is the associated Laguerre polynomial of indices $|\ell|$ and $p$. The overall factor ${\cal E}_0$ is a normalisation constant which is determined in terms of the applied power $\cal P$, evaluated as the integral of the z-component of the Poynting vector over the beam cross-section.  We have
\beq
{\cal P}=\frac{1}{2\mu_0}\int_0^{2\pi}d\phi\int_0^{\infty}|({\bf E}^*\times{\bf B})_z|\rho d\rho=\left(\frac{\pi\omega^2\epsilon_0 cw_0^2}{4}\right){\cal E}_0^2
\label{ee0}
\eeq
from which we can obtain ${\cal E}_0$ in terms of the power ${\cal P}$.

We first illustrate how the higher order affects radially-polarised Laguerre-Gaussian modes identified from the general formalism by setting $\cos{(\Theta_P)}=0$, so the helicity density is given by Eq.(\ref{emm}).   Figure 2 displays the helicity density for the case $m=4$ along with that of the first order ($m=1$) for $\ell=+1$ and $\ell=+2$.  
It is clear that for $\ell=+1, m=4$ the helicity density is super-chiral.  The chirality density here is always negative for $\ell=+|\ell|$ and is concentrated on the core at $\rho=0$. The case $\ell=+2, m=4$ is also super-chiral, but concentrated off-axis $\rho>0$.

Consider now the full higher order helicity desnity Eq.(\ref{heldens3}) for the particular case $\ell=1$ and the order is $m=10$, for illustration, and we choose $\cos{(\Theta_P)}=+1$, corresponding to right-circular polarisation.
The variations of the helicity density with $\rho/w_0$ for the case where $\ell=+1, +2$, are shown in Fig. 3.  We find, as in Fig. 2,  that for $\ell=+1$ the helicity density does not vanish at  $\rho=0$, in contrast to the case $\ell\geq 2$ where it always does.

The behaviour in the case $\ell=1$ can be explained by inspecting the general form of the helicity density terms which appear in Eq.(\ref{emm}) and also on the m-dependent terms in Eq.(\ref{heldens3}).  When applied to the Laguerre-Gaussian ${\cal F}$ for $\ell=1$, we have from Eq.(\ref{efftilde})
$
|{\tilde {\cal F}}_{\ell=1}|^2\propto \rho^2 
$
 and also we have
$
[{\tilde {\cal F}}'{\tilde {\cal F}}]_{\ell=1}\propto \rho.  
$
Once substituted in the relevant terms in the helicity density we see that the factor $1/\rho^2$ in the first term cancels the factor $\rho^2$ in the numerator and the $1/\rho$ in the second term cancels with the factor $\rho$ in the numerator.  The overall variation amounts to a non-zero value of the helicity at $\rho=0$ only in the case $\ell=1$.  This variation contrasts with the case $\ell\geq 2$ in which the numerators in the two terms have higher powers of $\rho$, guaranteeing that  the helicity density vanishes at $\rho=0$. 

Since the higher order helicity for $m>2|\ell|$ is dominated by the m-dependent terms we can evaluate the total integral of the helicity density due to the m-dependent terms in  Eq.(\ref{heldens3}) over the $x-y$ plane.   
First we note that the radial integral of all terms in the form $\frac{{\tilde {\cal F}'}{{\tilde{\cal F}}}}{\rho}$ are identically zero for all mode functions which satisfy ${\tilde {\cal F}}_{\{\ell\}}(0)=0={\tilde {\cal F}}_{\{\ell\}}(\infty)$.  We then have, for any ${\tilde {\cal F}}_{\{\ell\}}$, the helicity per unit length is 
\beq
{\cal {\bar C}}_{m}
=\frac{\epsilon_0 c^2}{4\omega}[m^2\cos{(\Theta_P)}-2m\ell]\int_{0}^{\infty}\rho d\rho\left[\frac{1}{\rho^2}|{\tilde {\cal F}}_{\{\ell\}}|^2\right]
\label{ceebar1}
\eeq
\begin{figure}
\includegraphics[width=1\linewidth]{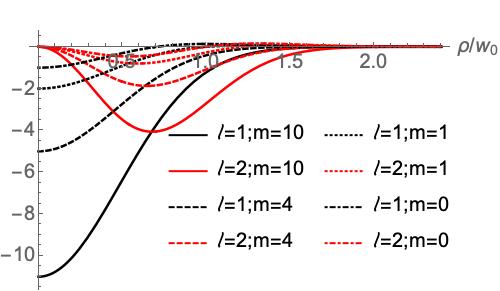}
\caption{Variations with $\rho/w_0$  of the helicity density, Eq.(\ref{emm}), due to modes of orders $m=0,1,4,10$. The plots concern radially-polarised Laguerre-Gaussians for which $\cos{(\Theta_P)}=0$ and the winding numbers are $\ell=+1$ and $\ell=+2$.  
When compared with the $m=0$ and $m=1$ plots we see that for $\ell=+1, m=4$ and $\ell=+1, m=10$ the helicity density in each case is super-chiral.  It is negative and concentrated on the core at $\rho=0$. Also by comparison, the $\ell=+2, m=4$ and $\ell=+2, m=10$ higher order modes are also super-chiral, but concentrated off-axis $\rho>0$.}
\end{figure}
\begin{figure}
\includegraphics[width=1\linewidth]{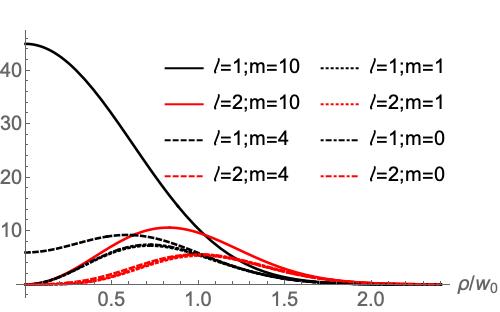}
\caption{Variations with $\rho/w_0$  of the full helicity density, Eq.(\ref{heldens3}), due to modes of order $m=0,1,4,10$. The plots concern circularly-polarised Laguerre-Gaussians for which $\cos{(\Theta_P)}=1$ and the winding numbers are $\ell=+1$ and $\ell=+2$.  
When compared with the zero-order $m=0$ plot we see that for $\ell=+1, m=4$ and $\ell=+1, m=10$ the higher order helicity density is strongly super-chiral and is concentrated on the core at $\rho=0$. Also by comparison, the $\ell=+2, m=4, 10$ are also super-chiral, but concentrated off-axis $\rho>0$.}
\end{figure}
 Substituting from Eq.(\ref{efftilde}) and using the integration variable $x=2\rho^2/w_0^2$ we have for the radial integral in Eq.(\ref{ceebar1})
\beq
{\cal I}=\frac{p!}{2(p+|\ell|)!}\int_0^{\infty} x^{|\ell|-1}e^{-x}[L^{|\ell|}_p(x)]^2dx\nonumber\\
=\frac{1}{2|\ell|}\label{ceebar}
\eeq
We finally obtain
\beq 
{\cal {\bar C}}_{m}={\cal L}_0\left[\frac{m^2\cos{(\Theta_P)}-2m\ell}{|\ell|}\right]\left(\frac{1}{k_z^2w_0^2}\right)
\label{final}
\eeq
where ${\cal L}_0={\cal P}/(k_zc^2)$ is a constant for a fixed power ${\cal P}$ and we have substituted for ${\cal E}_0$ using Eq.(\ref{ee0}). It is easy to check that ${\cal {\bar C}}_{m}$ has the dimensions of angular momentum per unit length.  Note that although the factor $1/k_z^2w_0^2$ in Eq.(\ref{final}) is typically small for $w_0^2>>1/k_z^2$, the higher order helicity for which $m\gg 2|\ell|$ would ensure super-chirality for relatively large $w_0\gg\lambda/2\pi$. Also we note from Fig. 3 in which $\Theta_P=0$ that two of the curves are identical, namely the one for the $m=4, \ell=+2$ case which is seen to have the same helicity curve as for $m=0, \ell=2$.  This can be verified from Eq.(\ref{heldens3}) but directly from Eq.(\ref{final}), both indicating that the $m$-dependent terms of the helicity vanish for $m=2\ell$ with $\ell>0$ and we are left with the $m$-independent chirality.  Also it can be seen that for $m<2\ell$ with $\ell>0$ we have negative contributions to the helicity from the $m$-dependent terms. Super-chirality arises when $m\gg 2|\ell|$. We have verified by direct analysis that the energy density behaves in the same manner as the helicity density as its expression contains similar terms to those entering the helicity density.

In conclusion, we have evaluated the chirality/helicity  densities for general paraxial light modes in which the state of polarisation is specified by a general point $(\Theta_P,\Phi_P)$ on the surface of the order $m$ Poincar{\'e} unit sphere, where $m$ is a positive integer. The general results obtained encompass a wide range of scenarios governed by their dependence on the Poincare sphere angles, the winding number $\ell$, the mode amplitude function and the higher order $m$.  In particular, for points  $(\pi/2,\Phi_P)$ on the equatorial circle, the helicity/chirality is found to be proportional to $m$, which means that the higher order modes exhibit super-chirality since it is enhanced $m$-fold relative to the helicity of an ordinary (order $m=1$) radially polarised mode. For all other points on the surface of the  Poincar\'{e} sphere the helicity is enhanced further by terms proportional to  $m^2$.  
We have also shown that a higher order $m$ Laguerre-Gaussian mode for which $\ell=+1$ is a strongly super-chiral vortex beam which is dominated by  the vortex core at $\rho=0$ and the helicity  at the core increases with increasing $m$. We have found that other higher order Laguerre-Gaussian modes for which $\ell>+1$ have off-axis maximum helicity  which is also super-chiral. These results strongly indicate the existence of a highly desirable super-chirality property of the higher order modes which, we suggest, is now ripe for direct experimental investigation.  There are diverse applications that can be envisaged, including improved interactions with chiral matter and stronger trapping and manipulation using optical spanners and tweezers, for example in micro-fluidics and improved encoding schemes for higher bandwidth optical communications. 
\bibliography{bibliography.bib}
\end{document}